# Spectroscopic evidence for possible quantum spin liquid behavior in a two-dimensional Mott insulator

Haiyang Chen[1]†, Fo-Hong Wang[1]†, Qiang Gao[1], Xue-Jian Gao[2], Zhenhua Chen[3], Yaobo Huang[3], Kam Tuen Law[2], Xiao Yan Xu[1,4#], Peng Chen[1#]

[1]Key Laboratory of Artificial Structures and Quantum Control (Ministry of Education), Shanghai Center for Complex Physics, School of Physics and Astronomy, Shanghai Jiao Tong University, Shanghai 200240, China.

[2]Department of Physics, Hong Kong University of Science and Technology, Clear Water Bay, Hong Kong, China.

[3]Shanghai Synchrotron Radiation Facility, Shanghai Advanced Research Institute, Chinese Academy of Sciences, 201204 Shanghai, China.

[4]Hefei National Laboratory, Hefei 230088, China

†These authors contributed equally to this work.

#Email: xiaoyanxu@sjtu.edu.cn;

pchen229@sjtu.edu.cn

Abstract

Mott insulators with localized magnetic moments will exhibit a quantum spin liquid (QSL) state when the quantum fluctuations are strong enough to suppress the ordering of the spins. Such an entangled state will give rise to collective excitations, in which spin and charge information are carried separately. Our angle-resolved photoemission spectroscopy (ARPES) measurements on single-layer 1T-$TaS_2$ show a flat band around the zone center and a gap opening of about 200 meV in the low temperature, indicating 2D Mott insulating nature in the system. This flat band is dispersionless in momentum space but shows anomalously broad width around the zone center and the spectral weight decays rapidly as momentum increases. The observation is described as a spectral continuum from electron fractionalization, corroborated by a low energy effective model. The intensity of the flat band is reduced by surface doping with magnetic adatoms and the gap is closing, a result from the interaction between spin impurities coupled with spinons and the chargons, which gives rise to a charge redistribution. Doping with nonmagnetic impurities behaves differently as the chemical potential shift dominates. These findings provide insight into the QSL states of strongly correlated electrons on 2D triangular lattices.

One of the most exotic phenomena in QSLs is the fractionalized excitations resulting in a separation of the spin and charge components of electrons, known as spinons and chargons [1-3]. Identification of these two quasiparticles is challenging as the former is charge neutral and carries spin and the latter is spinless. By removing an electron from a QSL, the hole state will separate into a spinon hole and a holon. In the mean-field approximation, the convolution of their dispersions will give rise to the electronic spectral function [Fig. 1(c)] [4], resulting in a composite spectrum and the signature should be readily probed by ARPES. As an example, the two-branch structure in 1D antiferromagnetic (AFM) chains has been shown as a result of spin-charge separation [5, 6].

1T-TaS$_2$ has been proposed to be a gapless QSL candidate [2, 3]. Experimental evidences include no long-range magnetic order down to 20 mK, QSL dynamical behavior revealed by nuclear quadrupole resonance, and finite residual linear term in the thermal conductivity [7-10]. The bulk system undergoes a first-order transition into a commensurate charge density wave (CCDW) state at ~180 K [11, 12]. In this phase, thirteen in-plane Ta atoms distorted into a Star of David (SOD) cluster to form a ($\sqrt{13} \times \sqrt{13}$) superstructure with one unpaired electron localized on the center Ta atom [Fig. 1(a)] [13]. A Mott insulator model is therefore natural to explain the ground state. However, the interlayer coupling effect from various stacking orders complicates the situation and the nature of insulating phase for the bulk material in the low temperature is still under debate [14-17].

In this work, we report a study of the band structure of single-layer 1T phase TaS$_2$ by ARPES and present the evidence for the electron fractionalization in the 2D triangular lattice corroborated by the theoretical analysis. The reduced dimension is ideal to explore the 2D Mott physics and its electron correlation effect without interlayer coupling. Using molecular beam epitaxy (MBE), we have successfully grown high-quality single-layer TaS$_2$ films, with sharp diffraction patterns measured by reflection high energy electron diffraction (RHEED), as shown in Fig. S1. A layer of 1T-TaS$_2$ is formed by a planar Ta triangular lattice sandwiched by two layers of S atoms. Similar to the bulk case, a triangular ($\sqrt{13} \times \sqrt{13}$) Star of David CDW superstructure was also observed in single-layer 1T-TaS$_2$ by scanning tunneling microscopy [18], but the band structure and temperature dependent behavior remain unexplored. Angle-integrated core-level scans [Fig. 1(d)] show the characteristic peaks of Ta. Splitting of the Ta 4$f$ states due to the inequivalent structures indicates a mixture of 1T and 1H phases [18-20]. The mixed phases are also evidenced in the ARPES bands [Fig. 1(e)].

Fig. 1(e) shows the ARPES spectra of single-layer TaS$_2$ taken along the $\overline{\Gamma M}$ direction at 10 K. A flat band labeled as α is located at ~-0.26 eV centered around the $\overline{\Gamma}$ point. This topmost band in 1T phase is almost dispersionless with a bandwidth (~10 meV), about 10 times less than the bulk case [14] and the system is insulating with a spectroscopic gap of 0.26 eV at 10 K (Fig. S4). It is consistent with the Mott insulator picture that decreased Coulomb interaction screening in the 2D limit gives rise to more localized orbitals and reduces the band width, as a consequence, the half-filled band crossing the Fermi level splits into the lower and upper Hubbard bands. Density functional theory (DFT) calculations predict a bandwidth of less than 30 meV for the topmost band (Fig. S7), in consistent with the experiments. The α' band is the folded band in the smaller CDW Brillouin zone caused by CDW superlattice potential. Below the flat band, a dispersive hole-like CDW folded band (β) is observed [Fig. 1(e) and Fig. S3], which disperses towards the flat band as the momentum increases. A feature with strong intensity denoted as the γ band is from Ta 5$d$ state. Two CDW gaps at binding energies of 0.3 and 0.9 eV develop at low temperatures and separate the γ band into several pieces. There is no obvious photon-energy dependence in these bands, in agreement with the 2D nature of the electronic structure of the system, as shown in Fig. S2. Note that the bands crossing the Fermi level are from the 1H-TaS$_2$ and do not interfere with the bands in the 1T phase. They can be used as a reference to roughly confirm the position of the Fermi level and determine the amount of chemical potential shift in doping experiments, as demonstrated in below. Two hole-like valence bands with binding energy at 1.3 eV centered about $\bar{\Gamma}$ are S-3*p*-derived bands and those at ~1.5 eV are from 1H states.

The metal-Mott insulator transition is investigated by taking systematic scans of the bands along $\overline{\Gamma M}$ (Fig. 2) as a function of temperature. As the lower Hubbard band becomes much blurry at higher temperatures and the energy of band top is difficult to be extracted, the leading-edge midpoint (LEM) of the energy distribution curves (EDCs) at the $\bar{\Gamma}$ point is used to illustrate the gap formation in 1T phase TaS$_2$. This LEM gap is expected to be related to that from the transport as this gap includes the tail of the lower Hubbard band which affects the transport measurements [21]. The Mott gap is diminishing at higher temperatures but is not fully closed at 330 K. The square of the extracted energy gap is plotted as a function of temperature in Fig. 2(c). As the commensurate CDW state and Mott phase are deeply coupled to each other, fitting to a semi-phenomenological BCS mean-field equation [22] [red curve in Fig. 2(c)] yields a CCDW-Mott transition temperature of $T_C$ = 353 ± 12 K. This single layer transition temperature is twice the bulk transition temperature (180 K), illustrating the robustness of the CCDW-Mott gap in 2D 1T-TaS$_2$. No obvious difference shown in the temperature dependent gaps between the cooling and heating processes indicates the hysteresis observed in the bulk [11, 12] is possibly from the competing stacking orders in the system.

With increasing temperature, the intensity of the flat band becomes much weaker in the high temperatures and this band nearly vanishes at 330 K. Interestingly, the gap between the α and α' bands become larger as the temperature increases, different from the prediction of the conventional band theory that a single and continuous band forms around the Fermi level along the $\overline{\Gamma M}$ direction (Fig. S7) [23, 24]. This result demonstrates the origin of the gap between α and α' bands is different from the CDW gaps along the γ band which are closing with increasing temperature. Other unusual properties of α band include large broadening in energy and fast decay of spectral weight as momentum increases. Although the α band is dispersionless in energy (~10 meV), the full width at half maximum (FWHM) of the EDC at the $\overline{\Gamma}$ point for α band is ~0.2-0.3 eV depending on the photon energy used, 2-3 times of that for 1H band and 1T band at a binding energy of 1.3 eV. Note that FWHM for γ band is also broad as there is overlap of multiple bands (Fig. S7). These features are not typical for the CDW bearing 2D materials and Mott insulators with symmetry-breaking orders [25-28].

These unconventional observations can be naturally explained by a spectral continuum from electron fractionalization in a QSL. Single-layer 1T-TaS$_2$ has been proposed as a gapless U(1) QSL candidate with a spinon Fermi surface (SFS) [3]. In the slave rotor formalism, the low energy dynamics of a U(1) QSL with SFS can be described by a mean field Hamiltonian, where spinons and chargons have their own dispersions [29]. A hole excitation fractionalizes into a spinon and a holon, leading to photoemission spectral continuum. As shown in Fig. 3, the model qualitatively captures the broad continuum feature near the $\overline{\Gamma}$ point and the suppressed spectral weight between the α and α' bands, supporting its origin from electron fractionalization. According to the effective model, the suppression of spectral weight between the α and α' bands results from the existence of the spinon Fermi surface. The Fermi wavevector of this spinon Fermi surface lies between the α and α' bands. With increasing temperature, the spectral weight of the continuum structure spreads out in both the model and the experiment. In the model, the Mott gap is set to 200 meV based on experimental data. The DFT-calculated bandwidth of the flat band is used for the holon dispersion and the spinon hopping isdetermined by matching the FWHM of the spectral continuum obtained in ARPES. Based on these considerations, the values for holon and spinon hopping are set to 3 meV and 30 meV, respectively. Note that a 30 meV spinon hopping is also suggested by Ref. 4. Such a large spinon hopping is consistent with the very large specific heat-temperature coefficients found in experiments [7]. Different values of holon and spinon hopping will result in different spectral shapes, e.g. differences in the relative strengths of $t_f$ and $t_X$ can give rise to the exchange of the upper and lower bounds (Fig. S11).

Charge neutrality makes spinon Fermi surface difficult to be directly observed. A promising way to detect it is to measure the excitations around the Hubbard band edges with magnetic doping as itinerant spinons will couple to the magnetic impurities, which forms the spinon Kondo cloud [4, 30]. Such a localized cloud can attract chargons and induce resonance peaks at the Hubbard band edges. To explore the effects of magnetic carriers on the spinons, we have employed surface doping of Fe on the single-layer 1T-TaS$_2$. As shown in Fig. 4, (a) and (b), with the sample at 10 K, increasing amounts of Fe doping causes the reduced intensity of the flat band and the closing of the CDW gaps. It is clear that there are some spectral intensity filling up the Mott gap around the lower Hubbard band edge with doping. This doping effect cannot be simply described by a rigid shift of the bands, another evidence showing that the single-layer 1T-TaS$_2$ is not a band insulator in the low temperature. The normalized EDCs at the zone center for different doping level are shown in Fig. 4(e). The band edge shifts towards the Fermi level at the high doping levels indicating the diminishing of the energy gap. To quantify how the energy gap evolves with the doping level, we extract the leading-edge midpoint of the EDCs [Fig. 4(g)]. The amount of dopant carrier density is obtained from the Luttinger area of the Fermi surface [31], in which the 1H phase bands dominates. At a dopant carrier density of ~2.5 x $10^{14}$ /cm$^2$, the gap is 15 meV smaller than the undoped system. A similar effect is also observed with Co adatoms on the surface of 1T-TaS$_2$ (Fig. S5).

The above observed reduced gap is a sign that there are some in-gap states filling up the energy gap around the lower Hubbard band edge. We did not see the distinct resonance peaks since they only emerge when the magnetic impurities are on the center of SOD [30]. As ARPES probes spatial-averaged electronic structure of samples, the bands with various Fe locations smear out the spectra. As a comparison, nonmagnetic impurity gives rise to a quite different result (Fig. 4 and Fig. S6). Fig. 4(c) shows the APRES spectra taken with Na atoms deposited on the single-layer 1T-TaS$_2$. The lower Hubbard band moves further away from the Fermi level and the spectral intensity becomes stronger with increasing amount of Na dosage. These findings are consistent with those reported in K doped bulk 1T-TaS$_2$ [32], in which, the half-filled band is argued to be saturated by the electron from the adsorbed K. Similar effect is also observed in Au doped single-layer 1T-TaSe$_2$ [30].

Note that the electron doping will in general result in a chemical potential shift of the samples. As is evidenced in two S 3$p$ bands at ~-1.3 eV, they shift towards higher binding energy with increasing doping levels, which we use as a reference for the shift of the chemical potential. By subtracting the value of chemical potential shift [Fig 4(h)], the energy position of the flat band does not vary much with Na doping, suggesting the chemical potential shift plays a major role in nonmagnetic impurity doping. However, the flat band shifts closer to the Fermi level with Fe, indicating the strong coupling effect between a QSL and magnetic adatoms.

Our ARPES study provides evidence supporting the existence of the QSL state in single-layer 1T-TaS$_2$. The low energy spectra reveal a spinon-holon continuum which can be qualitatively described by electron fractionalization in a U(1) QSL with SFS. The doping dependent results indicate the coupling between spin impurities-spinons and the chargons which gives rise to a charge redistribution and a reduced gap. The findings extend the phenomenon of spin-charge separation in 1D AFM chains to a 2D system and advance the understanding of the physics of QSL states in 2D triangular lattices.

**Acknowledgments**

We thank Prof. Yi Chen, Prof. Mingpu Qin for helpful discussions. The work at Shanghai Jiao Tong University is supported by the National Natural Science Foundation of China (Grant No. 12374188 and No. 12274289), the Ministry of Science and Technology of China under Grant No. 2021YFE0194100, No. 2022YFA1402400, No. 2022YFA1402702, and No. 2021YFA1401400, the Science and Technology Commission of Shanghai Municipality under Grant No. 21JC1403000, Shanghai Pujiang Program under Grant No. 21PJ1407200, the Innovation Program for Quantum Science and Technology (Grant no. 2021ZD0301902). P. C. and X. Y. X. thank the sponsorship from Yangyang Development Fund.

FIG. 1. Illustration of the electron fractionalization in the QSL state and band structure of single-layer 1T-TaS$_2$. (a) A schematic diagram of excitations in the QSL state. Entangled pairs are indicated by light blue ovals which cover two SODs in the triangular lattice. Blue arrows and red dashed circles correspond to spinons and holons, respectively. Green dots represent Ta atoms. (b) Schematic of the electron fractionalization process and spectral dispersion after convolution in single-layer 1T-TaS$_2$. Blue and purple solid curves represent the spinon and holon branches, respectively. Blue shadow area indicates the spinon-holon continuum. (c) Brillouin zones of the (1 × 1) and ($\sqrt{13} \times \sqrt{13}$) structures are outlined in blue and red, respectively. The red arrows indicate wave vectors of CDW. (d) A core-level photoemission spectrum taken with 90 eV photons for single-layer TaS$_2$ films. The 1H + 1T sample shows mixed Ta core level signals. (e) ARPES maps taken along $\overline{\Gamma M}$ with 45 eV $p$ polarized light at 10 K. The data show a strong flat α band and gap opening around the zone center.

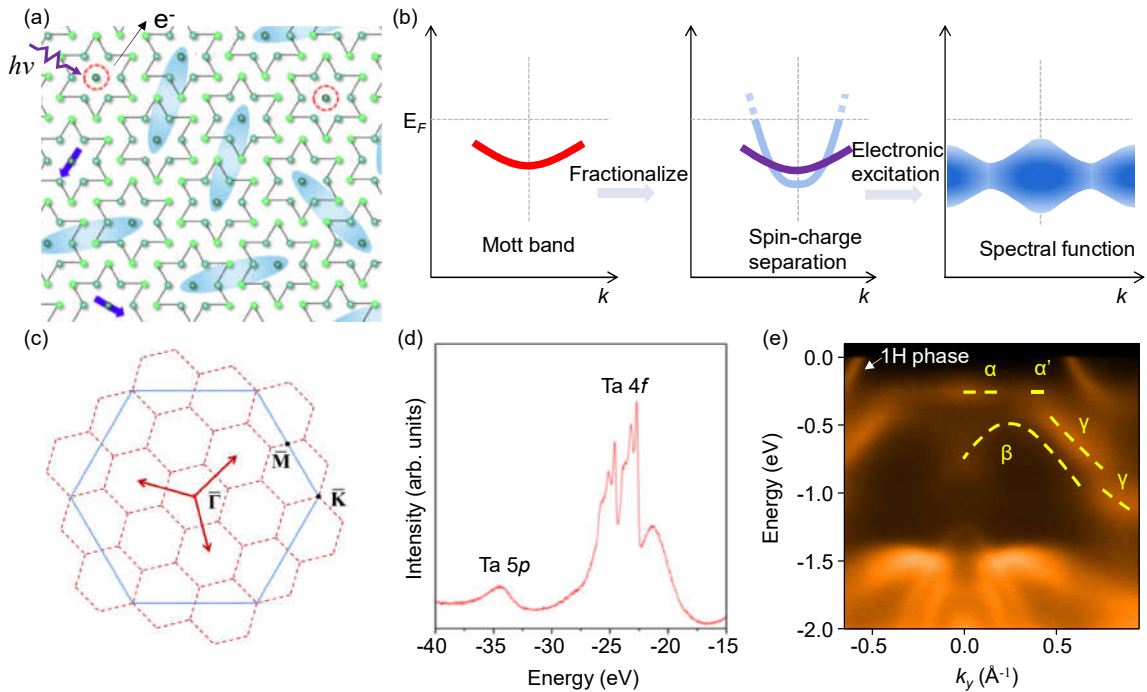

FIG. 2. Temperature dependence of the band structure and the energy gaps. (a) ARPES spectra along $\overline{\Gamma M}$ reveal the diminish of the flat band top and the bands shift towards the Fermi level when the temperature is increased from 10 to 330 K. (b) EDCs at the zone center at selected temperatures, the leading-edge midpoints are indicated by red and blue arrows for the EDCs at 10 and 330 K, respectively. (c) The extracted temperature dependence of the square of the energy gap. The red curves are fitting results using the BCS-type mean-field equation. Transition temperature $T_C$ is labeled. The error bar is deduced from the standard deviation of the fitting.

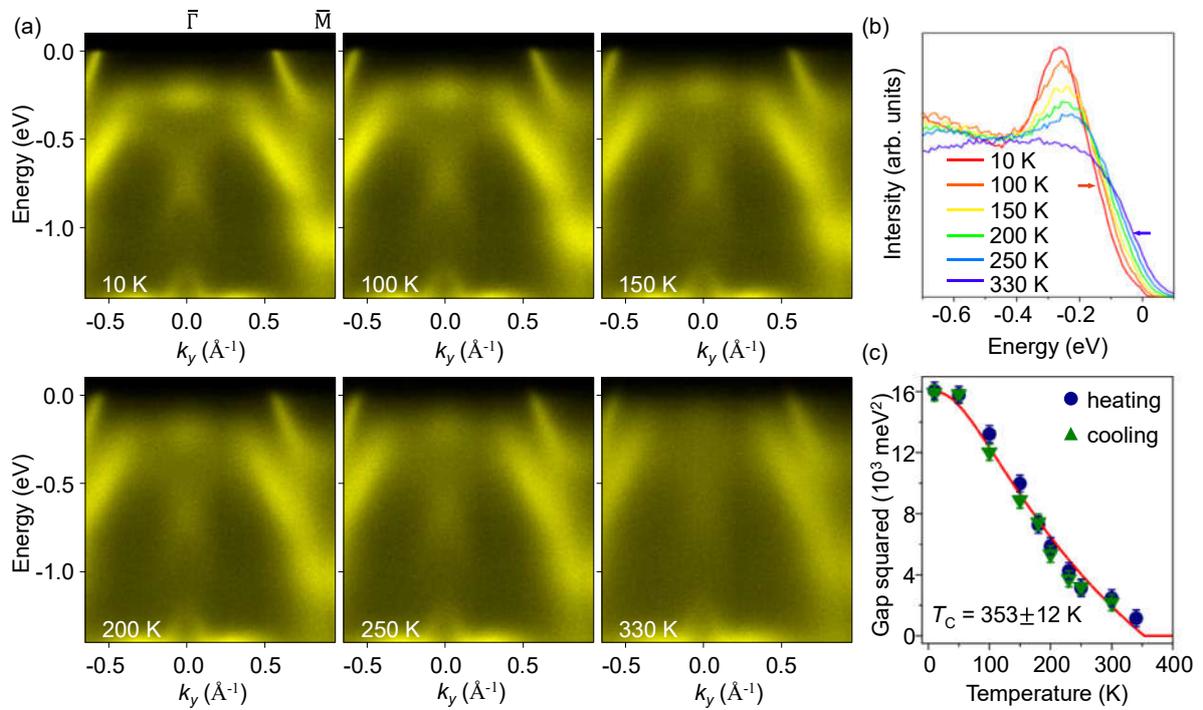

FIG. 3. Electronic structure and low energy effective model for U(1) QSL state. (a) ARPES spectra taken along the $\overline{\Gamma M}$ direction at 10 and 330 K. (b) Spectral function simulation results based on a low energy effective model at 10 and 330 K.

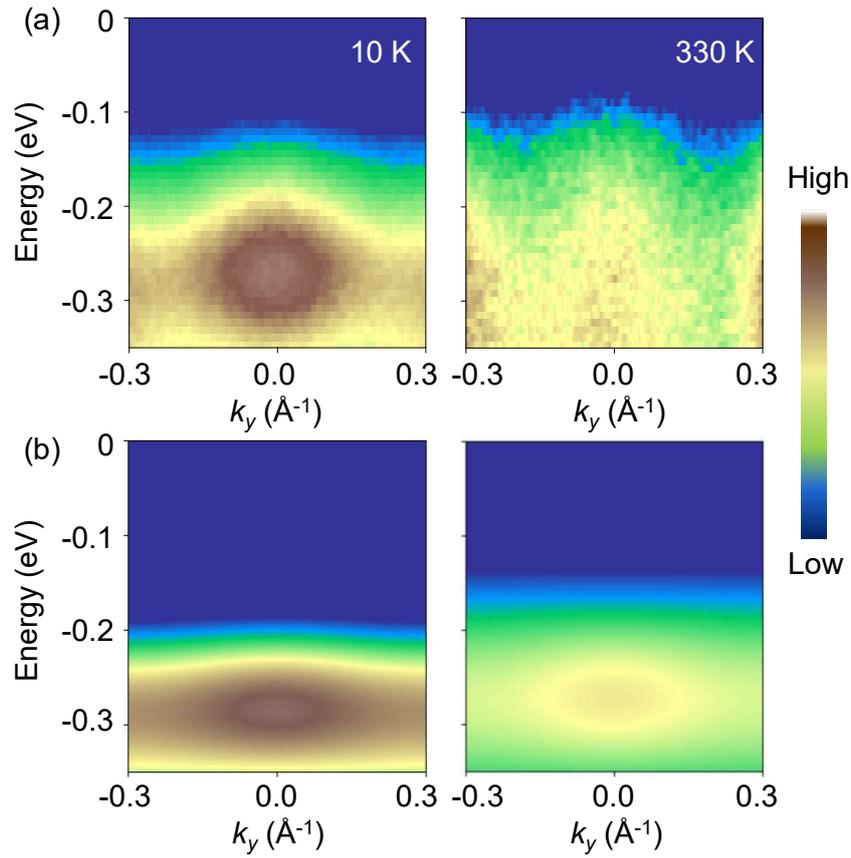

FIG. 4. Effect of surface doping on the band structure of 1T-TaS$_2$. Measured ARPES spectra taken along the $\overline{\Gamma M}$ direction at 10 K with (a) Fe adatoms and (c) Na adatoms, respectively. Surface electron density (*N*) determined from the Luttinger area of the Fermi surface is shown in each panel with a unit of $10^{14}$ cm$^{-2}$. (b) and (d) Corresponding symmetrized map to show the evolution of the flat band around the zone center with doping. (e) and (f) Normalized EDCs at the zone center with different doping levels. (g) The extracted doping dependence of the binding energy of the flat band around the zone center determined by the leading-edge midpoints, showing the flat band moves closer to the Fermi level at higher Fe doping levels but exhibits an opposite behavior with Na doping. (h) The binding energy of the flat band as a function of surface carrier density with a correction of chemical potential shift introduced by doping.

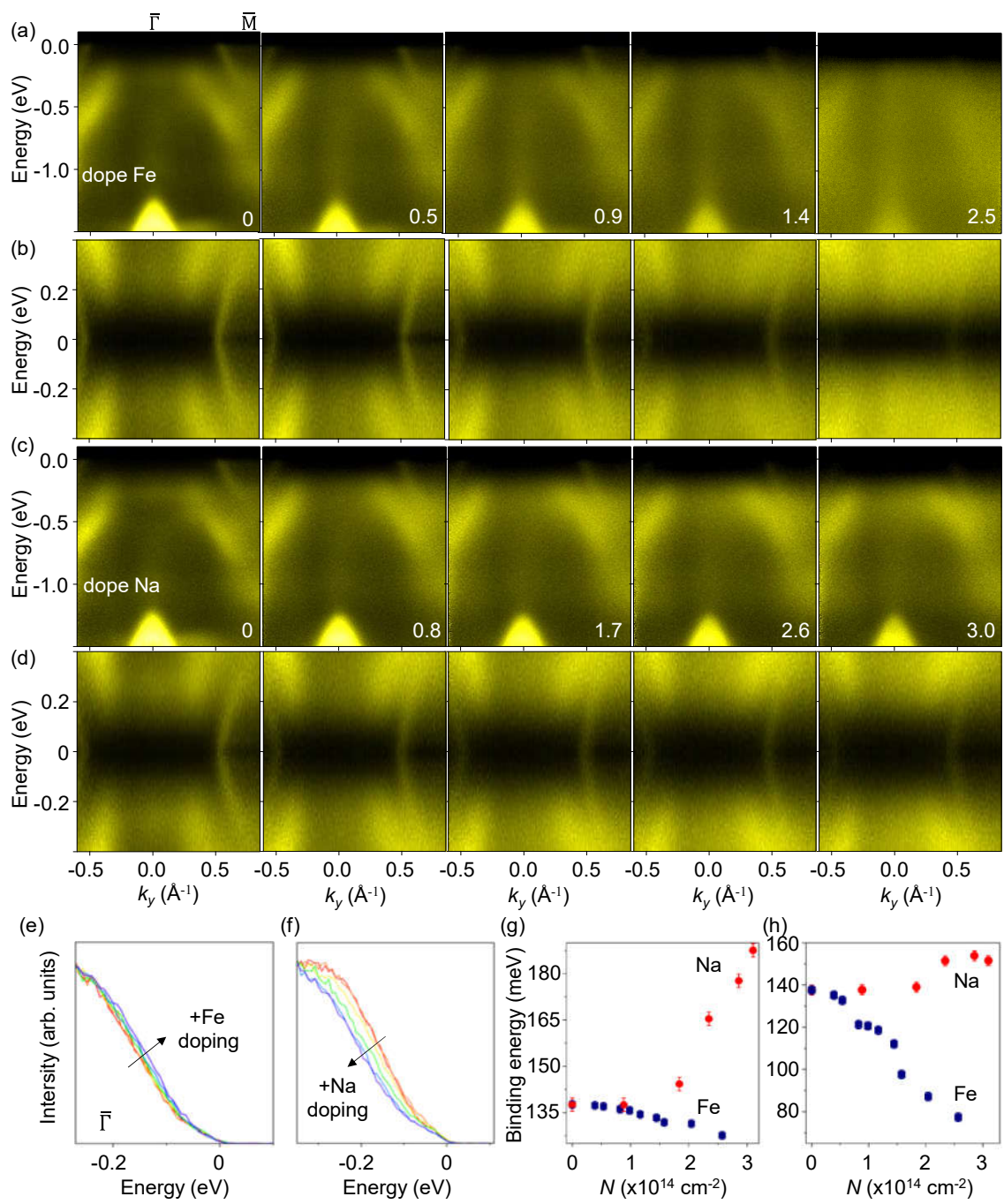